\documentclass{emulateapj}
\usepackage{apjfonts}
\lefthead{LEE ET AL.} 
\righthead{PLANETS IN BINARIES}

\newcommand{\zetavec}{\mbox{\boldmath $\zeta$}}

\newcommand{\thetae}{\theta_{\rm E}}

\begin{document}
\title{Microlensing Detections of Planets in Binary Stellar Systems}

\author{
Dong-Wook Lee\altaffilmark{1},
Chung-Uk Lee\altaffilmark{1},
Byeong-Gon Park\altaffilmark{1},
Sun-Ju Chung\altaffilmark{2},
Young-Soo Kim\altaffilmark{1},
Ho-Il Kim\altaffilmark{1},
Cheongho Han\altaffilmark{2, 3}
}
\altaffiltext{1}{Korea Astronomy and Space Science Institute, Hwaam-Dong,
Yuseong-Gu, Daejeon 305-348, Korea}
\altaffiltext{2}{Institute for Basic Science Research, Program of Brain 
Korea 21, Department of Physics, Chungbuk National University, Chongju 
361-763, Korea}
\altaffiltext{3}{Corresponding author}



\begin{abstract}
We demonstrate that microlensing can be used for detecting  planets 
in binary stellar systems.  This is possible because in the 
geometry of planetary binary systems where the planet orbits one of 
the binary component and the other binary star is located at a large 
distance, both planet and secondary companion produce perturbations 
at a common region around the planet-hosting binary star and thus 
the signatures of both planet and binary companion can be detected 
in the light curves of high-magnification lensing events.  We find 
that identifying planets in binary systems is optimized when the 
secondary is located in a certain range which depends on the type 
of the planet.  The proposed method can detect planets with masses 
down to one tenth of the Jupiter mass in binaries with separations 
$\lesssim 100$ AU.  These ranges of planet mass and binary separation 
are not covered by other methods and thus microlensing would be able 
to make the planetary binary sample richer.
\end{abstract}


\section{Introduction}

A binary star system is the most common result of star formation
process.  As the majority of stars belong to double or multiple 
star systems \citep{duquennoy91, eggenberger04a}, it is important 
to study the frequency of planets in binary systems and the properties 
of these planets for better understanding of the process of planet 
formation and evolution.

Planets in binaries have been discovered mainly through two channels.  
The first one is to perform dedicated surveys looking for planets 
in known visual or spectroscopic binaries \citep{konacki05, desidera06, 
eggenberger06}.  The second approach is to study the binarity of 
the hosts of planets discovered in planet surveys \citep{patience02, 
eggenberger04b}.  We now know $\sim 40$ planets in binaries or 
multiple systems.\footnote{See \citet{haghighipour06} for an 
up-to-date list of planets in binary systems.}  These systems are 
mostly wide binaries with separations of several hundred to several 
thousand AU.

However, the sample of planets in binaries is not still large enough 
for the statistical analysis of their properties. This is because 
binaries with separations $\lesssim 100$ AU are difficult targets 
for radial velocity surveys, which is the most productive technique 
among those currently being used for planet searches, and thus were 
often rejected from the samples.  In addition, due to the limitations 
of the available observational techniques, most detected objects are 
giant (Jupiter-like) planets, implying that the existence of smaller 
mass planets in multiple star systems is still an open question.

In this paper, we demonstrate that microlensing technique can be 
used for detecting planets in binary systems, especially for 
low-mass planets in binaries with separations $\lesssim 70$ AU.  
In the geometry of planets in binary systems where the planet orbits 
one component of a binary and the other binary star is located at 
a large distance, perturbations induced by the planet and binary 
companion can occur in the same region around the planet-hosting 
binary component.  Then, the signatures of both planet and binary 
companion can be identified in the light curve of a lensing event 
produced by the source star's passage close to the star hosting the 
planet.

The paper is organized as follows.  In \S\ 2, we describe lensing 
properties in various cases of lens geometry, including binary, 
planetary, and triple lensing.  In \S\ 3, we describe the channels 
of lensing events from which  planets in binaries can be detected.
We illustrate signatures of planets in binary stellar systems and  
investigate the range of binary separations within which detections 
of planets are optimized.  We also discuss about possible complications 
in the interpretation of the signals.  We summarize and conclude 
in \S\ 4.

\begin{figure*}[ht]
\epsscale{0.92}
\caption{\label{fig:one}
Magnification patterns of triple lens systems composed of a 
planet and binary stars. The coordinates are centered at the 
position of the planet-hosting star ({\it primary}) and the 
$x$-axis is aligned with the line connecting the primary and 
the other binary star ({\it secondary}).  All lengths are in 
units of the Einstein radius corresponding to the mass of the 
primary star, $\theta_{\rm E,1}$.  The secondary is on the 
right side and the position angle of the planet measured from 
the primary-secondary axis is $60^\circ$.  The projected distances 
of the planet ($\hat{s}_{\rm p}$) and secondary ($\hat{s}_{\rm b}$) 
from the primary are marked in each panel.  Note that notations 
with the {\it hat} represent length scales normalized by 
$\theta_{\rm E,1}$.  Also marked are the mass ratios of the 
planet-primary ($q_{\rm p}$) and secondary-primary ($q_{\rm b}$) 
pairs.  Panels in each {\it column} show the magnification 
patterns of lens systems with a common planet but with different 
separations to the secondary star.  Panels in each {\it row} 
show the cases where the size of the planet-induced caustic 
$\Delta\xi_{\rm c,p}$ is two times smaller (upper row), 
equivalent (middle row), and two times larger (lower row) than 
the size of the secondary-induced caustic $\Delta\xi_{\rm c,b}$, 
respectively.  Brighter greyscale represents the region of higher 
magnifications.  The figures drawn in soild curves are the caustics.  
The straight lines with arrows represent source trajectories and 
the light curves of the resulting events are presented in the 
corresponding panels of Figure~\ref{fig:three} (solid curves).
}\end{figure*}

\section{Multiple Lensing}

The lensing behavior of a planet in a binary stellar system requires 
the formalism of a triple lens with two equivalent-mass components 
and a very low-mass third body.  If a source star is gravitationally 
lensed by a lens system composed of $N$ point-masses, the equation 
of lens mapping from the lens plane to the source plane (lens equation) 
is expressed as \citep{witt90} 
\begin{equation}
\zeta = z - \sum_{j=1}^N {m_j/M \over \bar{z}-\bar{z}_{{\rm L},j}},
\label{eq1}
\end{equation}
where $\zeta=\xi+i\eta$, $z_{{\rm L},j}=x_{{\rm L},j}+ iy_{{\rm L},j}$, 
and $z=x+iy$ are the complex notations of the source, lens, and image 
positions, respectively, $\bar{z}$ denotes the complex conjugate of 
$z$, $m_j$ are the masses of the individual lens components, and 
$M=\sum_j m_j$ is the total mass of the system.  Here all lengths 
are normalized to the Einstein radius corresponding to the total 
mass of the lens system, i.e.
\begin{equation}
\thetae = \left[ {4GM\over c^2} 
\left( {1\over D_{\rm L}} - {1\over D_{\rm S}}  \right)
\right]^{1/2},
\label{eq2}
\end{equation}  
where $D_{\rm L}$ and $D_{\rm S}$ are the distances to the lens and 
source, respectively.   Finding the locations of images for a given 
positions of the lens and source requires inversion of the lens 
equation.  The lensing process conserves the source surface brightness 
and thus the magnifications $A_i$ of the individual images correspond 
to the ratios between the areas of the images and source.  For each 
image located at $z_i$, this ratio corresponds to the Jacobian of 
the lens equation, i.e.  
\begin{equation}
A_i = \left\vert \left( 1-{\partial\zeta\over\partial\bar{z}}
{\overline{\partial\zeta}\over\partial\bar{z}} \right)_{z=z_i}^{-1} 
\right\vert.
\label{eq3}
\end{equation}
For Galactic microlensing events, the typical separation between 
images is of the order of 0.1 mill-arcsec and thus the individual 
images cannot be resolved.  However, events can be noticed by 
the variation of the source star flux \citep{paczynski86}, where 
the magnification corresponds to the sum of the magnifications of 
the individual images, i.e.\ $A=\sum_i A_i$.

For a single-lens case, the lens equation is simply inverted to 
solve the image positions.  This yields two images located at 
\begin{equation}
{\bf u}_{{\rm I},\pm}={1\over 2}\left[ u \pm (u^2 +4)^{1/2}\right]
{{\bf u} \over u}, 
\label{eq4}
\end{equation}
where ${\bf u}=\zetavec-{\bf z}_{\rm L}$ is the separation 
vector between the lens and source.  The magnifications of the 
individual images are
\begin{equation}
A_\pm= {1\over 2} \left[ {u^2+2\over u(u^2+4)^{1/2}} \pm 1 \right],
\label{eq5}
\end{equation}
yielding a total magnification of
\begin{equation}
A=A_{+} + A_{-}={u^2+2 \over u(u^2+4)^{1/2}}.
\label{eq6}
\end{equation}

If a lens system is composed of more than two masses, the lens 
equation cannot be inverted algebraically.  One way to investigate 
the lensing optics for a multiple-lens system is expressing the 
lens equation as a polynomial in $z$ and finding the image 
positions by numerically solving the polynomial \citep{witt95}.
The advantage of this method is that it allows semi-analytic 
description of the lensing behavior and saves computation 
time.  However, the order of the polynomial increases as $N^2+1$
and thus solving the polynomial becomes difficult as the number 
of lenses increases.  In this case, one can still obtain the 
magnification patterns by using the inverse ray-shooting technique 
\citep{schneider86, kayser86, wambsganss90}.  In this method, a 
large number of light rays are uniformly shot backwards from the 
observer plane through the lens plane and then collected (binned) 
in the source plane.  Then, the magnification pattern is obtained 
by the ratio of the surface brightness (i.e., the number of rays 
per unit area) on the source plane to that on the observer plane.  
Once the magnification pattern is constructed, the light curve 
resulting from a particular source trajectory corresponds to the 
one-dimensional cut through the constructed magnification pattern.  
Although this methods requires a large amount of computation time
for the construction of the detailed magnification pattern, it 
has the advantage that the lensing behavior can be investigated
regardless of the number of lenses.

\begin{figure*}[ht]
\epsscale{0.92}
\caption{\label{fig:two}
Magnification patterns of lens systems composed of binary stars
but without planets.  The notations are same as those in 
Figure~\ref{fig:one}.  The light curves resulting from the source 
trajectories marked in the individual panels are presented in the 
corresponding panels of Figure~\ref{fig:three} (dotted curves).
}\end{figure*}

The number of images, their locations, and the resulting magnification 
pattern vary greatly depending on the number of lenses, their individual 
mass fractions, and the geometry of the lens system.  A multiple-lens 
system has a maximum of $N^2+1$ images and a minimum of $N+1$ images.  
One common important characteristic of multiple lensing is the formation 
of caustics.  Caustics represent the set of source positions at which 
the magnification of a point source becomes infinite.  For a binary-lens 
case, the caustic form a single or multiple sets of closed curves each 
of which is composed of concave curves (fold caustics) that meet at 
points (cusps).  Unlike the caustics of binary lensing, caustics of 
multiple lensing can exhibit self-intersecting and nesting.  The number 
of images changes by a multiple of two as the source crosses a caustic 
\citep{rhie97}.

\section{Signatures of Planet and Binary Companion}

\subsection{Perturbation Approach}

We consider the geometry of planets in binary stellar systems where 
the planet is orbiting one of the star in the binary and the other 
binary star is located at a larger distance.\footnote{Some planets 
are known to orbit around close binaries such as the planet orbiting 
the pulsar binary PSR 1620-26 in the globular cluster M4 
\citep{sigurdsson93}.  The microlensing method is inefficient in 
detecting planets in such systems and thus we do not consider this 
geometry.} Hereafter we refer the star hosting the planet as 
{\it primary} and the other binary star as {\it secondary}.  In this 
lens geometry, the resulting lensing behavior can be approximated as 
the superposition of those of the binary lens pairs composed of the 
primary-secondary stars and primary-planet.  This is possible because 
the lensing effects caused by the secondary star and the planet in 
the region around the primary star are small and thus can be treated 
as perturbations.  For the primary-secondary pair, the effect of the 
secondary is small because of the large distance to the secondary star 
\citep{dominik99}.  For the primary-planet pair, on the other hand, 
the effect of the planet is small because of the small mass ratio of 
the planet \citep{bozza99}.

\begin{figure*}[ht]
\epsscale{0.92}
\plotone{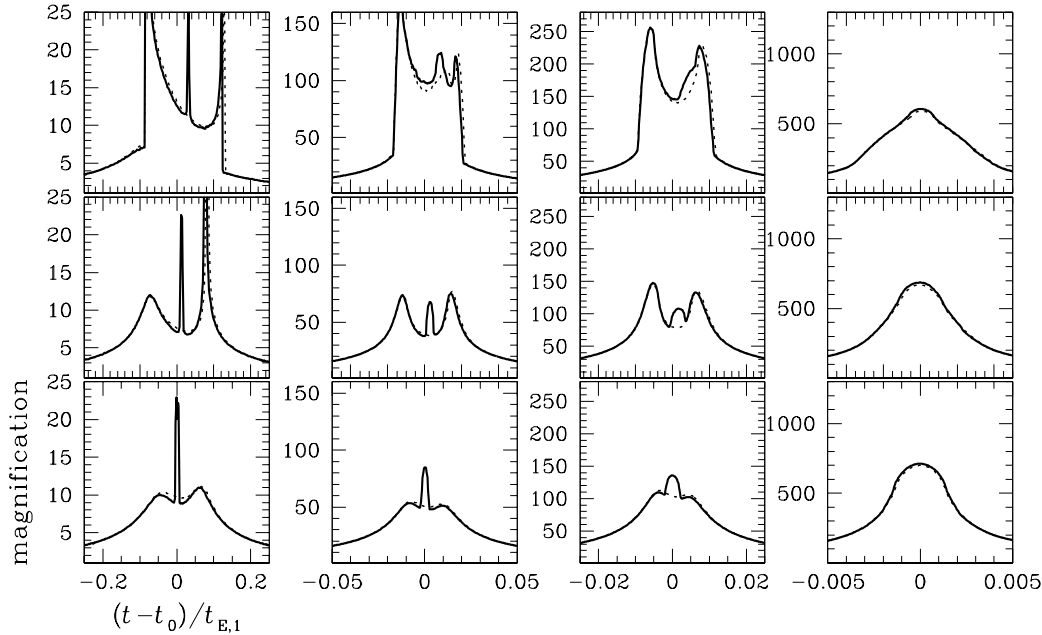}
\caption{\label{fig:three}
Light curves of events resulting from the source star trajectories 
marked on the magnification pattern maps in Figure~\ref{fig:one} and 
\ref{fig:two}.  Solid and dotted curves are the light curves with 
and without the planet, respectively.  For the construction of the 
light curves, finite-source effect is taken into consideration by 
assuming that the source star radius normalized by the Einstein 
radius is $\rho_\star=0.002$.  Time scale is normalized by the 
Einstein time scale corresponding to the mass of the primary.
}\end{figure*}

In the limiting case of a binary lens where the projected separation 
between the lens components is much larger than the Einstein radius, 
the lensing behavior in the vicinity of the primary star is approximated 
by the equation of the Chang-Refsdal lensing \citep{chang79, chang84, 
dominik99}, i.e.
\begin{equation}
\hat{\zeta} = \hat{z} - {1\over \hat{z}} + \gamma \hat{z}.
\label{eq7}
\end{equation}
Here the notations with the `{\it hat}' represent length scales 
normalized by the Einstein radius corresponding to the mass of 
the primary star, $\theta_{{\rm E},1}=\theta_{\rm E}[m_1/(m_1+
m_2)]^{1/2}$, $m_1$ and $m_2$ are the masses of the primary and 
companion stars, respectively, and $\gamma$ represents the shear 
induced by the secondary star.  The shear is related to the 
binary parameters by
\begin{equation}
\gamma=
{q_{\rm b}\over \hat{s}_{\rm b}^2};
\qquad q_{\rm b}={m_2\over m_1},
\label{eq8}
\end{equation}
where $\hat{s}_{\rm b}$ is the separation between the binary stars 
normalized by $\theta_{\rm E,1}$.  The validity of equation~(\ref{eq7}) 
implies that if the binary separation is sufficiently wide ($s_{\rm b}
\gg 1.0$), the lensing behavior in the region around the primary star 
can be approximated as that of a single point-mass lens superposed on 
a uniform background shear $\gamma$.  The result of the external shear 
is manifested as the formation of a small caustic around the primary 
with the shape of a hypocycloid with four cusps.  The width of the 
caustic as measured by the separation between the two cusps located 
on the binary axis is 
\begin{equation}
\Delta \xi_{\rm c,b} \simeq 
{4\gamma\over (1-\gamma)^{1/2}}=
{4q_{\rm b}\over \hat{s}_{\rm b}^2} \left( 1+{q_{\rm b}
\over 2\hat{s}_{\rm b}^2}\right).
\label{eq9}
\end{equation}
The caustic is tiny for a wide-separation binary because its size 
shrinks as $\Delta \xi_{\rm c,b} \propto \hat{s}_{\rm b}^{-2}$.  
However, a perturbation can occur if the source trajectory approaches 
close enough to the primary lens around which the caustic is located.  
During the time of perturbation, the source star is close to the 
primary lens and thus the magnification of the resulting event is 
very high.  Then, the signature of a wide-separation secondary is a 
short-duration perturbation near the peak of the light curve of a 
high-magnification event \citep{han03}.

Because of the small mass ratio of the planet, the light curve 
of a planetary lensing event is also well described by that of 
a single lens of the primary star for most of the event duration.  
For a planetary case, there exist two sets of disconnected 
caustics.  Among them, one is located away from the primary star.  
The other caustic, on the other hand, is located close to the 
primary lens.  The caustic located close to the primary (central 
caustic) has a wedge-like shape and its size as measured by the 
width along the star-planet axis \citep{chung05} is related to 
the planet parameters by
\begin{equation}
\Delta \xi_{\rm c,p} \simeq {4q_{\rm p} \over 
(s_{\rm p}-1/s_{\rm p})^2};\qquad 
q_{\rm p}={m_{\rm p}\over m_1},
\label{eq10}
\end{equation}
where $s_{\rm p}$ is the separation between the primary and planet 
measured in units of $\theta_{\rm E}$ and $m_{\rm p}$ is the mass 
of the planet.  We note that for the case of a planetary lensing 
where $q_{\rm p}\ll 1.0$, $\theta_{\rm E} \sim \theta_{\rm E,1}$ 
and thus $s_{\rm p}\sim \hat{s}_{\rm p}$.  The size of the caustics 
is maximized when the planet is located close to the Einstein ring 
of the primary star, $s_{\rm p}\sim 1.0$.  Since the central caustic 
is located close to the primary lens, the perturbation region around 
the central caustic induced by the planet overlaps with the perturbation 
region induced by the wide-separation binary companion.

\subsection{Optimal Lens Geometry}

Because the planet-induced perturbation region overlaps with the
perturbation region caused by a wide-separation secondary, the 
deviation induced by the planet can be additionally perturbed 
by  the binary companion.  This makes it possible to  use the 
microlensing technique as a tool to identify planets in binary 
systems.

To illustrate the feasibility of the microlensing detection of planets 
in binary stars, we present magnification patterns of triple lens 
systems composed of a planet and binary stars in Figure~\ref{fig:one}.  
In each map, the coordinates are centered at the position of the 
primary and the $x$-axis is aligned with the primary-secondary axis.  
Since the perturbation occurs near the peak of a seemingly single-lens 
event caused by the primary, we normalize all lengths in units of 
the Einstein radius corresponding to the mass of the primary star.  
The secondary is on the right side and the position angle of the 
planet measured from the primary-secondary axis is $60^\circ$.  
The separations to the planet ($\hat{s}_{\rm p}$) and secondary 
($\hat{s}_{\rm b}$) are marked in each panel.  Also marked are the 
mass ratios of the planet-primary ($q_{\rm p}$) and secondary-primary 
($q_{\rm b}$) pairs.  Panels in each {\it column} show the magnification 
patterns of lens systems with a common planet but with different 
projected distances to the secondary star.  Panels in each {\it row} 
show the cases where the size of the planet-induced caustic $\Delta
\xi_{\rm c,p}$ is two times smaller (upper row), equivalent (middle 
row), and two times larger (lower row) than the size of the 
secondary-induced caustic $\Delta\xi_{\rm c,b}$,  respectively.  
Brighter greyscale represents the region of higher magnifications.  
The figures drawn in solid curves represent the caustics.  Note that 
the caustic curves exhibit self-intersecting and nesting, that are 
the characteristics of multiple lensing.  In Figure~\ref{fig:two}, we 
also present maps of binary lenses without planet for the comparison 
of the patterns with the triple-lens systems.  The straight lines 
with arrows in the magnification pattern maps represent source 
trajectories and the light curves of the resulting events are presented 
in the corresponding panels of Figure~\ref{fig:three}.  Since both 
caustics induced by the planet and secondary companion are small, 
finite size of the source star would be important in lensing light 
curves.  We therefore consider the finite-source effect by assuming 
that the source star has a radius equivalent to the Sun, i.e. $r_\star
=1\ R_\odot$.  For a typical Galactic event caused by a low-mass 
stellar object of $m_1=0.3\ M_\odot$ and with distances to the lens 
and source stars of $D_{\rm L}=6$ kpc and $D_{\rm L}=8$ kpc, respectively, 
the angular Einstein radius is $\theta_{\rm E} =0.32$ mas and the 
radius of the source star normalized to the Einstein radius is 
$\rho_\star \sim 0.002$.  For other combinations of the lens and 
source parameters, the normalized source size is 
\begin{equation}
\rho_\star = 0.0009 \left( {r_\star\over R_\odot}\right)
\left( {0.3\ M_\odot\over M}\right)^{1/2}
\left( {D_{\rm L}\over 6\ {\rm kpc}}\right)^{1/2}
\left( 1-{D_{\rm L}\over D_{\rm S}}\right)^{-1/2}.
\label{eq11}
\end{equation}
For the construction of the magnification maps and light curves, 
we use the inverse ray-shooting technique due to the difficulty in 
solving tenth-order triple-lens polynomial lens equation incorporating 
finite-source effect.

\begin{figure}[t]
\epsscale{1.2}
\plotone{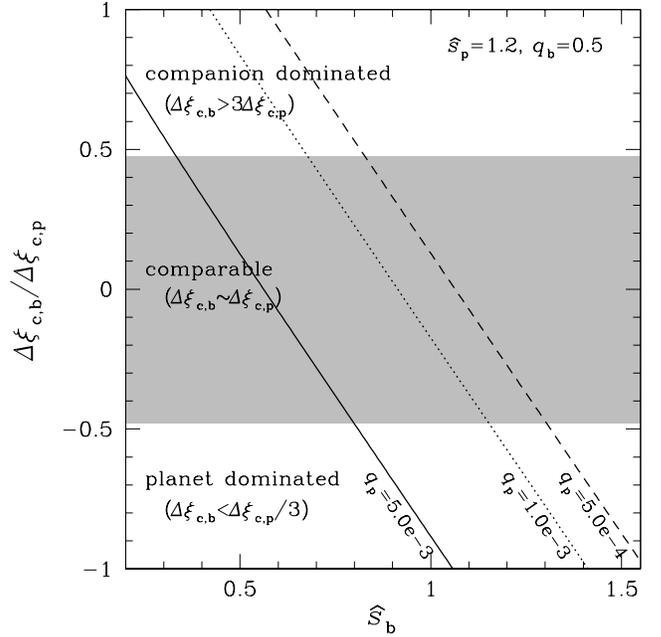}
\caption{\label{fig:four}
Size ratio between the caustics induced by the planet ($\Delta 
\xi_{\rm c,p}$) and secondary companion ($\Delta \xi_{\rm c,b}$)
in triple lens systems composed of a planet and binary stars as 
a function of the separation between the primary and secondary 
star.  The shaded area represents the region where the ratio is 
$1/3\leq \Delta \xi_{\rm c,b}/\Delta\xi_{\rm c,p} \leq 3$.  
The binary separation $\hat{s}_{\rm b}$ is normalized by the Einstein 
radius corresponding to the mass of the primary star.  To draw the 
curve, we adopt a distance to the planet of $\hat{s}_{\rm p}=1.2$ 
and a secondary/primary mass ratio of $q_{\rm b}=0.5$ as representative 
values.  Note that notations with {\it hat} represent length scales 
normalized to the Einstein radius corresponding to the mass of the 
planet-hosting star.
}\end{figure}

\begin{deluxetable}{ccc}
\tablecaption{Optimal Binary Separation\label{table:one}}
\tablewidth{0pt}
\tablehead{
\multicolumn{1}{c}{planet} &
\multicolumn{2}{c}{optimal range of secondary separation} \\
\multicolumn{1}{c}{mass ratio} &
\multicolumn{1}{c}{normalized unit} &
\multicolumn{1}{c}{physical unit} 
}
\startdata
$5.0\times 10^{-3}$  &  $2.1\lesssim \hat{s}_{\rm b}\lesssim 6.7$    & $4.2\ {\rm AU}\lesssim d_{\rm b}  \lesssim 13.4\ {\rm AU}$ \\
$1.0\times 10^{-3}$  &  $4.8\lesssim \hat{s}_{\rm b}\lesssim 14.2$   & $9.2\ {\rm AU}\lesssim d_{\rm b} \lesssim 28.4\ {\rm AU}$ \\
\smallskip
$5.0\times 10^{-4}$  &  $6.7\lesssim \hat{s}_{\rm b}\lesssim 20.1$   & $13.4\ {\rm AU}\lesssim d_{\rm b} \lesssim 65.2\ {\rm AU}$ 
\enddata 
\tablecomments{ 
Ranges of binary separation for optimal microlensing detections 
of planets in binaries.  Here $\hat{s}_{\rm b}$ represents the 
binary separation normalized by the Einstein radius corresponding 
to the mass of the planet-hosting star.  The physical separation 
$d_{\rm b}$ is determined by assuming that the physical Einstein 
radius is $r_{\rm E}=2.0$ AU.
}
\end{deluxetable}

From the figures, one finds that the magnification pattern of the 
binary star with a planet is different from either the primary-secondary 
or primary-planet pair and thus the resulting light curve can produce 
distinctive signatures in the lensing light curves.  One also finds 
that identifying planets in binary systems is optimized when the 
secondary is located in a certain range of separation from the primary.  
This optimal separation range varies depending on the type of the 
planet.  To produce noticeable planetary signature, planets should be 
located close to the Einstein ring of the primary star.  Under this 
lens geometry, if the secondary is located not far enough from the  
primary, the perturbation induced by the secondary dominates over the 
planet-induced perturbation.  If the secondary is located too far away 
from the primary, on the contrary, its signature would be too small 
to be noticed.  Then, identifying the signatures of both planet and 
secondary would be optimized when the companion is located at a 
separation where the amount of secondary-induced perturbation is 
equivalent to that of the planet-induced perturbation.  When the 
planet is located within this optimal range, we find that planets 
can be detected with mass ratios down to $q\sim 5\times 10^{-4}$, 
which corresponds to one tenth of the mass of the Jupiter.

In Figure~\ref{fig:four}, we present the size ratio between the 
caustics induced by the planet and secondary companion as a function 
of the binary separation.  The curves with different line types show 
the ratios for planets with different mass ratios.  To draw the curve, 
we adopt a planetary separation of $\hat{s}_{\rm p}=1.2$ and a binary 
mass ratio of $q_{\rm b}=0.5$ as representative values.  The shaded 
area represents the region where the sizes of the two caustics induced 
by the planet and secondary are comparable and thus the chance of 
detecting the signatures of both planet and secondary is relatively 
high.  In Table~\ref{table:one}, we also present the optimal range of 
binary separations both in normalized and physical units.  The physical 
separation is determined by assuming that the physical Einstein radius 
is $r_{\rm E}=D_{\rm L} \theta_{\rm E} =2.0$ AU.  We find that although 
varies depending on the planet type, the optimal range of the binary 
separation for planet detection is $\lesssim 70$ AU.  Binaries 
with separations in this range is not being covered by the current 
radial velocity method.

\subsection{Interpretation of Signatures}
 
Since the perturbations caused by the planet and the wide-separation
binary companion occur in a common region and at a similar location of 
the lensing light curve, one might question whether the light curve 
produced by a binary lens with a planet could be mimicked by that of 
a simple binary-lens or a single planetary event by appropriate 
modification of the lens parameters.  However, we note that although 
the perturbation regions induced by the planet and the companion are 
similar, they are not identical.  As a result, the characteristic 
shape of the perturbation such as the multiple-peak features shown 
in Figure~\ref{fig:three} cannot be produced by a single companion.

Another related question would be whether the light curve could be 
mimicked by that of an event caused by a single stellar lens with 
multiple planets \citep{gaudi98}.  In this case, the light curve can 
exhibit multiple-peak features.  However, the perturbations induced 
by planets in general have different characteristics from those 
induced by wide-separation companions and the two different types of 
perturbation can be well distinguished as demonstrated in practice 
for the case of the lensing event MACHO 99-BLG-47 \citep{albrow02}.  
In addition, the individual perturbations are in many cases well 
separated, allowing investigation of the individual perturbations 
\citep{han05}.  Therefore, it would be possible to distinguish the 
two possible degenerate cases.

\section{Conclusion}

We demonstrated that microlensing technique can be used for 
the detections of planets in binary stellar systems, especially 
for low-mass planets in binaries with small separations.  The 
signatures of both planet and binary companion can be detected 
in the light curve of a high-magnification lensing event.  
High-magnification events are the prime target of high-cadence 
follow-up microlensing observations currently being conducted to 
search for extrasolar planets \citep{abe04, cassan04, park04} 
and two planets were actually detected through this channel 
\citep{udalski05, gould06}.  We found that identifying planets 
in binary systems is optimized when the secondary is located in 
a certain range which depends on the type of the planet.  The 
proposed method can detect planets with masses down to one tenth 
of the Jupiter mass in binaries with binary separations $\lesssim 
100$ AU.  These ranges of planet mass and binary separation are 
not covered by other methods and thus microlensing would be able 
to make the planetary binary sample richer.

\acknowledgments 

D-WL, C-UL, Y-SK, and H-IK acknowledge the support from Korea 
Astronomy and Space Science Institute.  CH and B-GP are supported 
by the grant (KRF-2006-311-C00072) of the Korea Research Foundation.  
We would like to thank M. Dominik for making helpful comments
on the paper.

\end{document}